# Applying an XML Warehouse to Social Network Analysis

## Lessons from the WebStand Project

### Benjamin Nguyen

University of Versailles
benjamin.nguyen@prism.uvsq.fr

### Antoine Vion

University of Aix-Marseille II
antoine.vion@univmed.fr

### François-Xavier Dudouet

Université Paris-Dauphine
fxdudouet@dauphine.fr

### Loïc Saint-Ghislain

Ecole des Mines de Nancy
loic.saint-ghislain@mines.inpl-nancy.fr

*"History evidences it, sociology extracts it."*

*- Nadine Gordimer, South African Novelist*

## ABSTRACT

In this paper we present the state of advancement of the French ANR *WebStand* project. The objective of this project is to construct a customizable XML based warehouse platform to acquire, transform, analyze, store, query and export data from the web, in particular mailing lists, with the final intension of using this data to perform sociological studies focused on social groups of World Wide Web, with a specific emphasis on the temporal aspects of this data. We are currently using this system to analyze the standardization process of the W3C, through its social network of standard setters.

## 1.     INTRODUCTION

In this paper, we describe our platform, *WebStand[1]*, currently under development, to be used by sociologists when studying information found on the Web, and in particular analyzing social behavior on mailing lists, forums or any place in which (tracked) discussions take place on the Web. Our current focus is the analysis of the W3C standardization mechanism around the XQuery recommendation.

Indeed, Information Technology is only just receiving attention from sociologists, and our goal is to create new tools for sociologists to assess and analyze this domain.

Our approach, when designing our initial platform architecture, was to consider, in conjunction with sociologists what sort of information they whished to obtain, and what sort of analysis they wanted to run. A preliminary study led us to the following conclusions:

Traditionally, sociological data consist of reports, questions and interviews. On the contrary, in the Web context, the data manipulated is **electronic**: mailing lists, homepages, and institution or company pages. Our goal is to discover, extract, and analyze actors of this field, their positions, their relationships, and their influence, etc. All this data is particularly adapted to automatic processing.

The *WebStand* approach is based on the use of a semi-structured temporal XML content warehouse to store the data, and graphically generated XQueries to analyze it. Let us stress that our warehouse aims to cover the whole Extract Transform and Load (ETL) scope of a sociological application. Our goal in this short paper is to focus on the architecture and temporal model of our application, briefly present the modules already developed, and give some sociological results that illustrate the sort of information that we can calculate easily.

## 2.     ARCHITECTURE AND MODULES

The WebStand architecture is shown in Figure 1. WebStand is implemented in Java, and is running using the JDBC compatible MonetDB-XQuery [3] database. The modules developed include (a) a simple schema editor (b) an XML querying and visualization tool, geared towards mailing lists analysis, (c) a CV crawler and analyzer based on the Exalead crawler[2], (d) an email list crawler, extractor and cleaner, (e) a conversion module to export the data to external sociology applications. Current extensions of the system concern mainly improving the ergonomics of these modules and improve application tailored web data acquisition modules, that are currently rule based information extraction of pages retrieved by the exalead.com crawler.

Although we use MonetDB XQuery database to store the data, in some cases where the queries can not be run (such as queries using temporal functions) we use Saxon-B to compute the result.

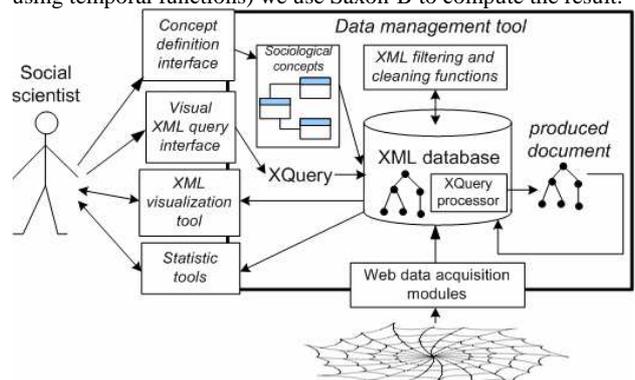

**Figure 1- WebStand Architecture**

The global use case is the following: a social scientist defines the concepts he is interested in, choosing from already existing concept (such as *person* or *email*) that can be extended with his own. This sociological model is (for the moment manually) translated into an XML Schema, used to store information extracted from the web by the acquisition modules. This XML

[1] This work is partially funded by the French ANR-JCJC-05 "WebStand".

Schema is also used to help the sociologist generate graphical queries, using a QBE-like interface, developed in our visualization and query tool. We used QBE rather than XQBE [1] due to the widespread use of Microsoft Access by the sociologists we work with, but we are considering alternative query interfaces based on XQBE. *WebStand* also provides simple XSL to export XML data in many formats used in the sociology world, although in a forseable future, we envision these applications to be all compatible with a simple XML format.

A preliminary study using our tool on 8 public mailing lists, related to XQuery and XML Schema has been performed. We are currently working on analyzing the data provided by all the public mailing lists of the W3C working groups.

## 3. SOCIOLOGICAL RESULTS

The corpus we focused on consist of 20.697 emails posted over the course of 4 years (from April 02 – to April 06) by about 3000 different "physical" people (i.e., after grouping emails together based on our heuristics, we identified 2923 different "entities"), analyzed according to activism on the lists and their participation in the writing of working drafts or recommendations. These emails originated from approximately 2000 different domains (Institutions or Internet Service Providers, our heuristics led us to 2076 different domains)

It is possible to run any query on this data, we show here simple aggregate results obtained to illustrate simple yet non the less valuable participation information.

Table 1 illustrates activism within W3C. It contains anonymized data showing the number of posts made by a single person: the top poster scored 1077 different posts. It is interesting to note that only 4 posters posted over 500 messages, and that only 500 posters out of 3000 posted over 5 messages. Turning to table 2, posts are now grouped by institution. We see that Microsoft and IBM dominate the playfield with Oracle tailing them. W3C posters are of course present. It is interesting to point out that the posts made by software AG all came from the same person, who went on to create his own XSL/XQuery company. Public research organizations such as universities are only represented by Edinburgh, UK, and although some public researchers post via their personal address (yahoo, aol, etc.) their participation is low, as show in Table 3, which illustrates the number of different posters, by domain name: 111 different people posting from yahoo.com posted 288 messages. On the other hand companies in terms of participation are once again IBM, Microsoft, and Oracle. We can see that Microsoft participant were extra-active, since only 20 people (compared to IBM's 35) posted nearly twice the number of emails. On the other end of the scope, universities and public research institutions are unable to mobilize a large number of active participants.

| | |
|---|---|
| 1. | 1077 |
| 2. | 730 |
| 3. | 683 |
| 4. | 604 |
| 5. | 423 |
| 6. | 385 |
| 7. | 373 |
| 8. | 318 |
| 9. | 225 |
| 10. | 223 |
| 11. | 207 |
| 12. | 203 |
| 13. | 198 |
| 14. | 197 |
| 15. | 169 |

**Table 1- Post count per person**

| | |
|---|---|
| microsoft.com | 1547 |
| ibm.com | 978 |
| softwareag.com | 681 |
| w3.org | 623 |
| oracle.com | 564 |
| cogsci.ed.ac.uk | 555 |
| acm.org | 485 |
| mhk.me.uk | 425 |
| nag.co.uk | 318 |
| yahoo.com | 288 |
| aol.com | 259 |
| datadirect.com | 212 |
| sun.com | 206 |
| arbortext.com | 203 |
| metalab.unc.edu | 196 |
| CraneSoftwrights.com | 180 |
| hotmail.com | 168 |
| kp.org | 165 |
| jclark.com | 141 |
| bea.com | 125 |

**Table 2- Post count per institution**

Table 4 shows the number of technical reports signed by members of institutions that signed at least one XQuery related[2] recommendation. Once again we see that IBM outnumbers

---

[2] We selected 28 technical reports in the recommendation process that appeared in the discussions on the list.

Microsoft by 2:1 both on the number of different authors and on the number of recommendations. Universities are also nearly non-existent. From a "neutral" sociologist point of view, these results point to the conclusion that corporations seem to dominate XQuery standard setting.

| | |
|---|---|
| yahoo.com | 111 |
| hotmail.com | 101 |
| w3.org | 99 |
| ibm.com | 35 |
| fake.invalid | 32 |
| excite.com | 27 |
| aol.com | 24 |
| microsoft.com | 20 |
| oracle.com | 20 |
| gmail.com | 18 |

**Table 3- Posters per domain**

| INSTITUTION | TYPE | # INDIV | REC. | W3C WG NOTES | DRAFTS |
|---|---|---|---|---|---|
| IBM | Corp | 11 | 8 | 2 | 3 |
| Oracle | Corp | 8 | 6 | 1 | 6 |
| AT&T | Corp | 2 | 4 | | 3 |
| Microsoft | Corp | 5 | 4 | | 2 |
| Unknown | n.a. | 2 | 3 | | |
| Sun Microsystems | Corp | 1 | 3 | | |
| Data Direct Technologies | Corp | 1 | 2 | 2 | 2 |
| University of Edimbourg | Uni | 2 | 2 | 1 | |
| Saxonica | Corp | 1 | 2 | | |
| Infonyte GmbH | Corp | 1 | 1 | 2 | |
| Brown University | Uni | 1 | 1 | | |
| CommerceOne | Corp | 1 | 1 | | |
| Inso | Corp | 1 | 1 | | |
| Kaiser Permanente | Org | 1 | 1 | | |
| SIAC | Corp | 1 | 1 | | |

**Table 4- Recommendation information**

Information used to create Table 4 was entered by hand using our temporal model detailed in section 4. We are currently in the process of automating authoring information from the versions (from WD to REC) of one W3C technical report found on the Web.

Other results that are produced by our system are *social graphs*, that indicate common participation on a thread, *answering*

*profiles* that indicate with which other list participants a given person privileges discussion, we can not provide them here due to the fact these graphs are place consuming, but we give one example in the Appendix, and we also refer to [4] for more examples of these graphs.

## 4. META MODEL AND QUERIES

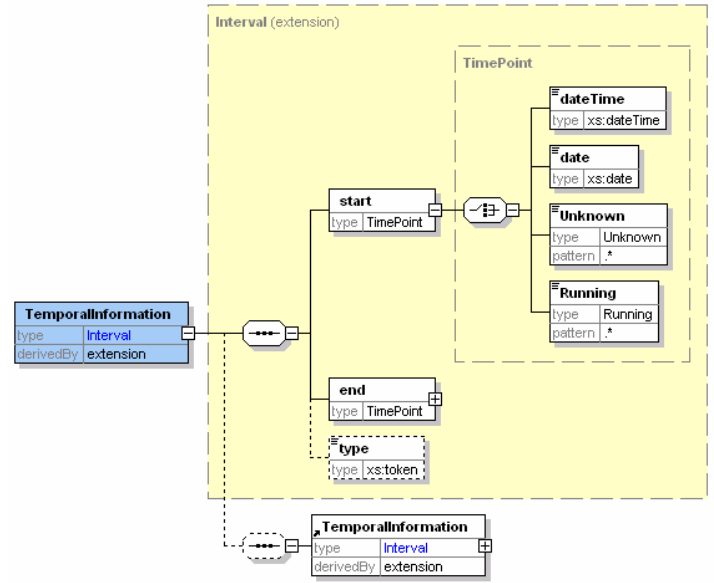

**Figure 2- Temporal XSD**

```
<person>
        <name>Doe</name>
        <firstname>John</firstname>
        <functions>
                <function>XMLCorp. CEO</function>
        </functions>
</person>
```

**Figure 3a- Non Temporal XML**

```
<person>
 <name>Doe</name>
 <firstname>John</firstname>
  <functions>
   <function>XML Corp. CEO
    <TemporalInformation>
     <start><date>2001-1-1</date></start>
     <end><date>2003-31-12</date></end>
     <type>valid</type>
    <TemporalInformation>
     <start><date>2004-6-6</date></start>
     <end><running></running></end>
     <type>valid</type>
    </TemporalInformation>
   </TemporalInformation>
  </function>
 </functions>
</person>
```

**Figure 3b-Temporalized XML information**

The data stored in the warehouse respects a given sociological schema that is generated by our tool (i.e., for the moment, we support *person*, *institution*, and *email* schemas). This data has specific temporal aspects, due to its inherent sociological nature. First of all, any information in the database (e.g., John Doe is XML Corp. CEO) is temporal. We want to store information regarding the fact that John became CEO at a certain point in time $t_1$(2001-1-1), and changed position at time $t_2$(2003-31-12). Moreover, this information was given to us at time $t_3$(2004-6-6). To represent this, we add two TemporalInformation nodes to the data, as shown in Figure 3b. The *event* element conveys the semantics attached to the temporal interval. This allows our model to capture the traditional validity time or transaction time aspects [5], but also to be fully flexible (any type of custom event can be defined).

If we receive new and contradictory information, for instance we learn at date t4(2005-4-10) the fact that John left his job at t2'(2003-30-11), this is simply captured by adding a new temporal annotation to the function node, and annotating this annotation to indicate the validity time's validity, and changing the <end> value of the previous validity node (no example shown due to place limitations).

Temporal data is accessed using XQuery, and we are currently implementing temporal query plan optimizations over MonetDB, which does not currently support temporal data.
We are also extending our sociological XML model to include sourcing and quality (i.e., where did we find the information, and what degree of credit to give to it.)

Current work also involves automatically extracting technical reports temporal information, cross referenced with authors and their affiliation at the time the report was published, and storing it in XML using our temporal model.

## 5. CONCLUSION

In this short paper, we present a brief overview of the architecture and functionalities of the *WebStand* platform and give some brief results of a study of the W3C. For more details on the sociological results, we refer to [4]. Our current experience shows that use of XML and XQuery through simple graphical interfaces simplifies the accessibility of XQuery to novice users, such as sociologists. The flexibility of our temporal model has allowed us to capture all the data collection situations that we have encountered so far, and it is our belief that such software can be used in various other sociological applications to analyse behaviors.

# 7. APPENDIX

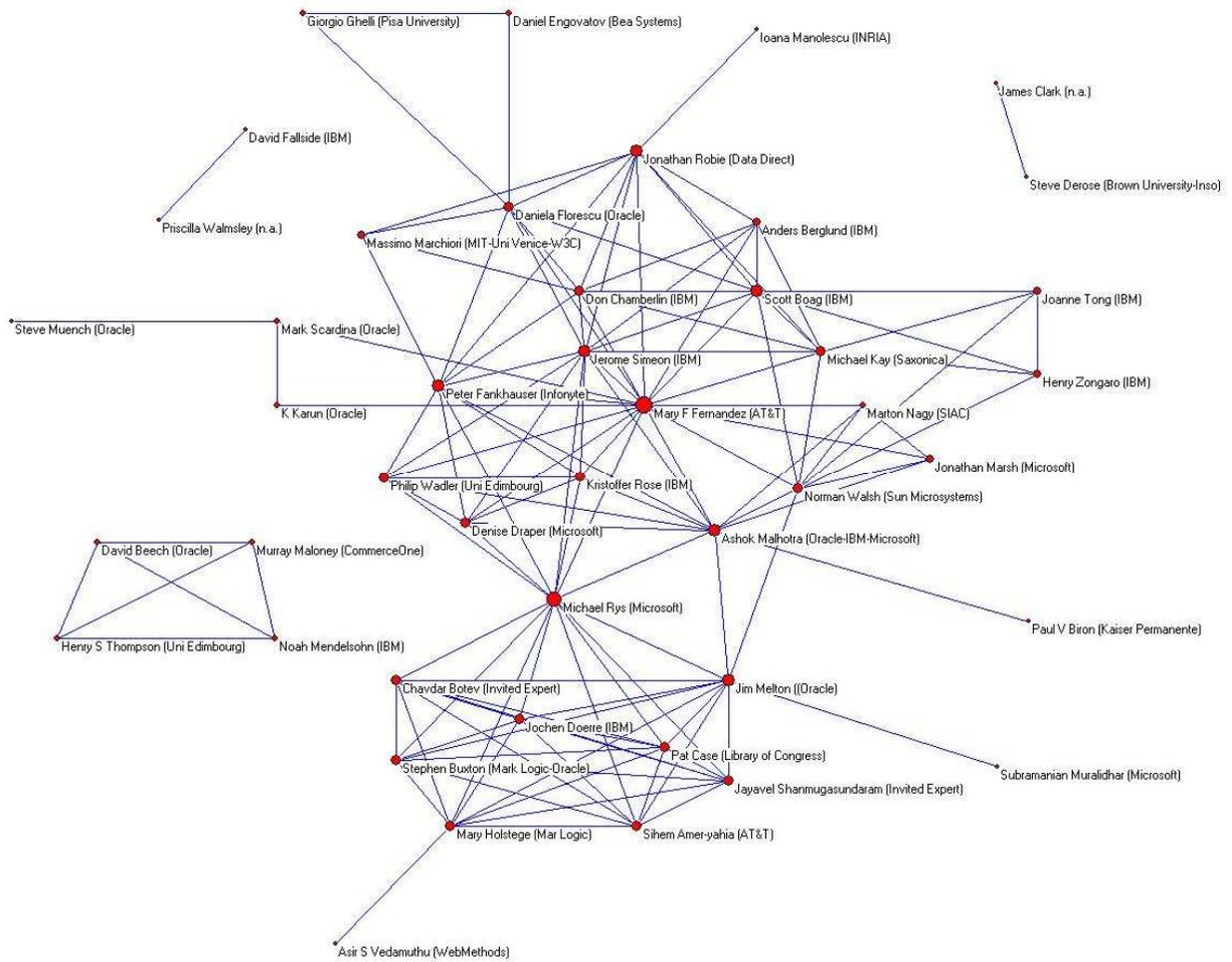

**Figure 3- Recommendations co-authors institutional mapping[3]**

---